# Two New Molecular Nitrogen Phases near Megabar Pressures

Alexander F. Goncharov ,[1] Elena Bykova ,[2] Iskander Batyrev,[3] Maxim Bykov,[4] Huawei Chen,[5] William Palfey,[1,5] Mahmood Mohammad,[5] Stella Chariton,[6] Vitali Prakapenka,[6] Jesse S. Smith [7]

[1] *The Earth and Planets Laboratory, Carnegie Institution for Science, Washington, District of Columbia 20015, USA*

[2] *Institute of Geosciences, Goethe University Frankfurt, Altenhöferallee 1, 60438 Frankfurt am Main, Germany*

[3] *U.S. Army DEVCOM Army Research Laboratory, FCDD-RLA-WA, Aberdeen Proving Ground, Maryland 21005, USA*

[4] *Institute of Inorganic and Analytical Chemistry, Goethe University Frankfurt, Max-von-Laue-Straße 7, 60438 Frankfurt am Main, Germany*

[5] *Department of Mathematics, Howard University, Washington, District of Columbia 20059, USA*

[6] *Center for Advanced Radiation Sources, University of Chicago, Lemont, IL 60637, USA*

[7] *HPCAT, X-ray Science Division, Argonne National Laboratory, Argonne, Illinois 60439, USA*

Molecular nitrogen shows extraordinary structural diversity near the polymeric transition, where multiple phases are metastable. We report two new molecular phases. The first, tζ-N$_2$, is a polytype of monoclinic *C*2/*c* ζ-N$_2$ with a tripled *c*-axis and 96 atoms per unit cell. The second, ξ-N$_2$, is a previously unknown hexagonal phase (*P*6*cc*) with 112 atoms per unit cell. Both were synthesized in a diamond anvil cell by laser heating ζ-N$_2$ to 1800-2500 K at 78–98 GPa. Structures were solved by single-crystal X-ray diffraction, confirmed by Raman spectroscopy, and supported by first-principles calculations. tζ-N$_2$ likely corresponds to the previously reported κ-N$_2$ phase.

Over the past decade, exploration of nitrogen's high-pressure phase diagram has advanced dramatically, driven by single-crystal (SC) X-ray diffraction (XRD) at pressures near and above 1 Mbar [1-3]. Earlier studies relied largely on Raman spectroscopy and powder XRD at ambient or elevated temperatures, using resistive or occasional laser heating [4-11]. SCXRD on well-prepared samples now enables precise determination of unit-cell parameters, symmetry, atomic positions, and disorder, allowing previously tentative phase assignments and phase boundaries to be established with far greater accuracy. This is particularly crucial for systems like nitrogen, with complex, metastable-rich phase diagrams.

SCXRD is especially critical for complex molecular phases with large unit cells, whose detection and structural solution are otherwise extremely challenging. Powder XRD and Raman spectroscopy often lack sufficient sensitivity and resolving power, and first-principles structure prediction is limited by unit-cell size and symmetry [12-14]. Identifying highly complex multimolecular phases therefore requires incorporating experimental constraints into structural searches, as recently demonstrated for ζ-$N_2$ [2] and ι-$N_2$ [1]. By contrast, the simpler θ-$N_2$ phase (8 atoms per unit cell) was theoretically predicted and recently confirmed experimentally by SCXRD and Raman spectroscopy [3].

Nitrogen exemplifies the transformation from molecular to polymeric bonding under compression, with numerous molecular phases and extensive metastability. Despite recent progress, many aspects of its high-pressure phase diagram remain unresolved, including the kinetics of phase transformations and the unexpectedly high polymerization pressure [3, 4, 6, 7, 10, 15-17]. Understanding the physical origins of this structural diversity is essential to elucidate the mechanisms governing metastability and polymerization.

We report the discovery of two previously unknown molecular phases of nitrogen, unexpected because all previously predicted phases had been identified and all experimentally reported structures determined—except κ-$N_2$ [5] observed above 120 GPa at 300 K and ζ'-$N_2$ [18], which forms on unloading amorphous η-$N_2$ at about 100 GPa. The first, a polytype of ζ-$N_2$ [2], shares its symmetry but has a unit cell tripled along *z*; theory suggests it is more stable than ζ-$N_2$ above 35 GPa and may correspond to the reported κ-$N_2$ [5]. The second has hexagonal symmetry *P*6*cc* and a host–guest type structure with molecular chains confined in cylindrical channels. Both are highly complex, with 96 and 112 atoms per unit cell, making prediction difficult.

The experimental procedure follows our previous studies on molecular crystals, including $CO_2$ [19], $CH_4$ [20], $I_2$ [21], and $N_2$ [3] (see also Supplementary Materials). Raman spectroscopy was performed on samples that were laser-heated in a diamond anvil cell (DAC). A key feature of this approach is the use of a metallic heat absorber, which enables efficient heating of molecular nitrogen at the pressures of interest. Heating prevents trapping in kinetically arrested phases that form during low-temperature compression and allows access to metastable phases otherwise unattainable. In this study, we used two less-common absorbers, Cu and Ag.

SCXRD was performed at beamlines 16-ID-B (HPCAT) and 13-ID-CD (GSECARS) of the Advanced Photon Source (APS), Argonne National Laboratory. Monochromatic X-ray beams were focused to 1–3 µm spots, with wavelengths of 0.34453 Å (HPCAT) and 0.3344 and 0.4133 Å (GSECARS). Diffraction was recorded using area detectors: PILATUS3 X 2M CdTe (HPCAT) and EIGER2 S CdTe 9M (GSECARS).

Experiments were conducted by gas-loading $N_2$ at 0.15 GPa into a DAC chamber containing <3 µm thick foils of Ag and Cu in a laser-drilled rhenium gasket. Symmetric DACs with ±32° openings were used, with pressures of 50–110 GPa monitored via the $v_1$ $N_2$ vibron of $\zeta$-$N_2$ as a marker [22] and, at the beamlines, in situ from powder XRD of fcc Cu and Ag using published equations of state [23, 24]. Samples were stepwise heated to 1300–3000 K determined spectro-radiometrically by near-infrared laser irradiation coupled to the metal flakes, with precise control of power and temperature [25]. At high temperatures, $N_2$ becomes fluid, enabling rapid reactions with the metal foils that produced new diffraction peaks assigned to polynitride phases; detailed characterization of these nitrides will be reported elsewhere.

The temperature was subsequently reduced stepwise over several minutes, during which nitrogen crystallization was observed in simultaneously collected powder XRD patterns. The quenched samples were mapped by powder XRD with 1–2 µm step sizes to identify regions showing multiple Bragg diffraction spots. SCXRD measurements were then performed at these locations following established procedures [3, 20, 21].

Heating Ag in an $N_2$ medium induces various phase transformations and chemical reactions, depending on the initial pressure, temperature, and heating duration (Table S1, Supplementary Materials). At 58 GPa, $\varepsilon$-$N_2$ transforms to $\zeta$-$N_2$ near the heated region, as determined by SCXRD measurements with structural parameters consistent with previous reports [2]. Gentle heating of the same sample at 78 GPa preserved $\zeta$-$N_2$, but SCXRD measurements at several positions near the heating spot revealed numerous additional reflections, which were assigned to a new $N_2$ phase, which we denote as $\xi$-$N_2$ (Fig. 1(a), Table S2, Supplementary Materials). No chemical reaction was detected at this stage based on XRD and Raman mapping near the heated region.

The new phase has hexagonal symmetry ($P6cc$) with $a = b = 14.105(3)$ Å and $c = 4.786(1)$ Å and contains 56 $N_2$ molecules per unit cell, exceeding the 48 molecules in the previously reported $\iota$-$N_2$ [1] and setting a new record for nitrogen structures. The structure is complex, with most molecules tilted relative to the $c$-axis, forming channels with two distinct molecular sites (Fig. 1(b)), where the molecules lie in the *ab* plane. Molecules at one site, located at the unit-cell vertices, exhibit pronounced disorder, modeled by distributing the occupancy among three positions with equal (1/3) occupancy.

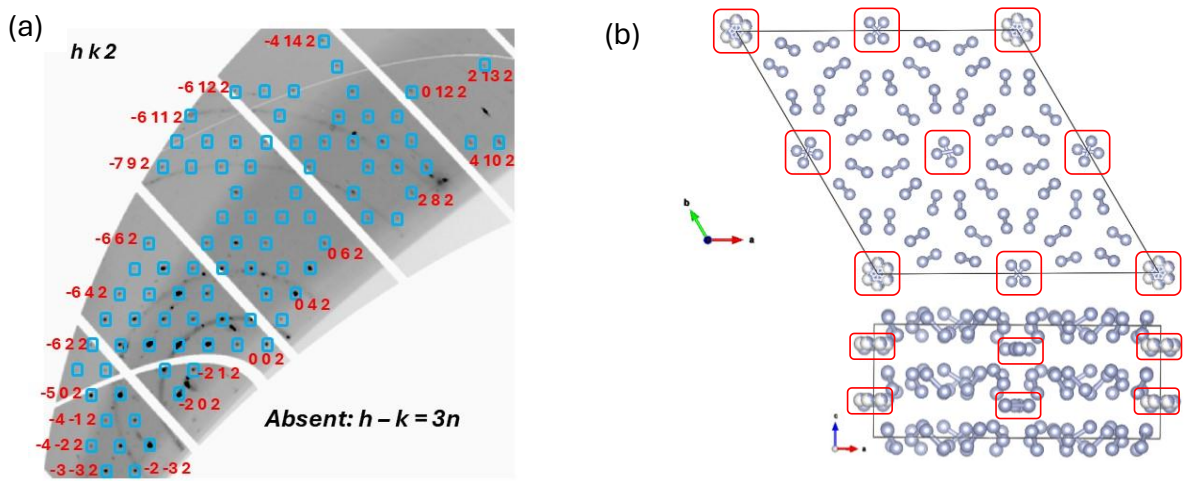

**Figure 1.** (a) Reconstructed reciprocal lattice planes of ξ-$N_2$ at 78 GPa showing hexagonal symmetry in the *xy* plane. Diffraction spots from the sample (blue box) were indexed and used to solve the *P6cc* $N_2$ (see Table S2 in the Supplemental Material). Systematic absences (extinction conditions) for this phase are indicated. (b) Crystal structure of ξ-$N_2$ projected onto the *xy* and *xz* planes. Molecules forming the central chains within the surrounding framework are highlighted (red box).

Prolonged laser heating to higher temperatures at 78 GPa, followed by slow cooling, resulted in the formation of $AgN_5$, similar to that reported in Ref. [26]; details will be presented elsewhere. At this stage, only ζ-$N_2$ was observed, while ξ-$N_2$ was no longer detected, likely having transformed back to ζ-$N_2$.

A second Ag–$N_2$ sample (Table S1, Supplementary Materials), prepared similarly to the first, was laser heated at 70 GPa and transformed to ι-$N_2$ in the heated region, while ζ-$N_2$ persisted in less-heated areas. Subsequent compression to 108 GPa followed by prolonged laser heating led to substantial formation of $AgN_5$, preventing determination of the nitrogen structure in the heated region by XRD. Raman spectroscopy indicates that ζ-$N_2$ is present in both heated and unheated regions of the high-pressure chamber.

In experiments with Cu foils in an $N_2$ medium, we focus on the run performed at 98 GPa. Prior to laser heating, ζ-$N_2$ was identified by Raman spectroscopy and powder XRD. Upon continuous increase of laser power, the radiatively measured temperature saturated near 1600 K, likely corresponding to melting of $N_2$. Further increase in laser power beyond a threshold caused a runaway temperature rise to ~3000 K. After quenching to room temperature, the sample consisted of a heterogeneous assemblage of several phases identified by XRD.

XRD mapping around the heated region revealed several distinct phases. In the central area, where the highest temperatures were reached, Cu and $N_2$ reacted completely to form new Cu nitrides with complex, large unit cells (Fig. S1, Supplemental Material), distinct from those reported at 52 GPa[27] and not predicted theoretically[28,29]. Their structures and compositions will be reported elsewhere. In the same region, we also identified a complex $N_2$ phase, whose structure was solved by single-crystal XRD (Fig. 2, Table S3, Supplemental Material). ζ-$N_2$ and ξ-$N_2$ were also detected in the quenched sample in less strongly heated regions.

The new molecular nitrogen phase adopts the same *C2/c* symmetry as ζ-$N_2$ [5, 9], whose structure was recently solved by SCXRD [2]. In contrast to ζ-$N_2$, however, the present phase contains three times as many molecules per unit cell (48), resulting from an unfolding along the *c* direction (Fig. 2). The enlarged unit cell arises from a modulation of the orientations of two types of molecules, which is absent in ζ-$N_2$. In the new phase, denoted tζ-$N_2$, the modulated molecules are canted with respect to the *ac* plane (Fig. 2), whereas in ζ-$N_2$ the corresponding molecules lie nearly within this plane. Smaller but similar modulations occur at the remaining molecular sites.

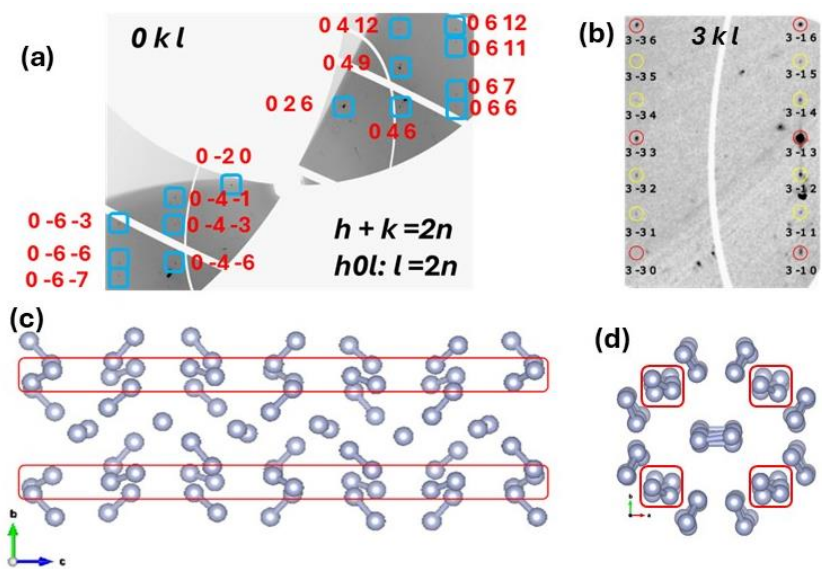

**Figure 2.** (a,b) Reconstructed reciprocal-lattice planes of tζ-$N_2$ at 98 GPa showing pseudohexagonal symmetry in the *xy* plane (a) and tripling of the unit cell along *c* (b). Diffraction spots from the sample (blue box in (a), circles in (b)) were indexed and used to solve the *C2/c* structure (see Table S3, Supplemental Material). Systematic absences are consistent with this space group; in (b), strong reflections with *l* = 3n and weaker reflections

with *l* = 3n + 1 and 3n + 2 demonstrate the tripled periodicity. **(c,d)** Crystal structure of tζ -$N_2$ projected onto the *xy* and *yz* planes. Molecules showing the largest orientational modulation along *c*, responsible for the unit-cell tripling, are highlighted (red box).

Raman spectroscopy was used to corroborate the XRD identification of the new phases by probing intramolecular vibrons and lattice (translational and librational) modes. Group-theoretical analysis and first-principles calculations (Supplemental Material) were performed to support the assignments. Raman spectra collected at 78 GPa in experiments on ξ-$N_2$ show two distinct responses (Fig. 3). In the unheated region, the spectra are characteristic of ζ-$N_2$, whereas in the heated region the signal is markedly different. The lattice modes of ξ-$N_2$ are broader and weaker than those of ζ-$N_2$. The high-frequency vibron modes of ξ-$N_2$ are also distinguishable from those of ζ-$N_2$ in their number, position, and linewidth.

The broadening of the lattice modes in ξ-$N_2$ is consistent with the large number of Raman-active modes predicted by group-theoretical analysis (Table S4, Supplemental Material) and first-principles calculations (Fig. 3). The experimental frequencies agree reasonably with theory when compared with the calculated Raman-intensity envelope. In the vibron region, quantitative agreement requires a uniform frequency shift of ~400 cm$^{-1}$, somewhat larger than previously reported [2]. This discrepancy likely reflects the harmonic approximation and limitations in the theoretical description of intramolecular interactions.

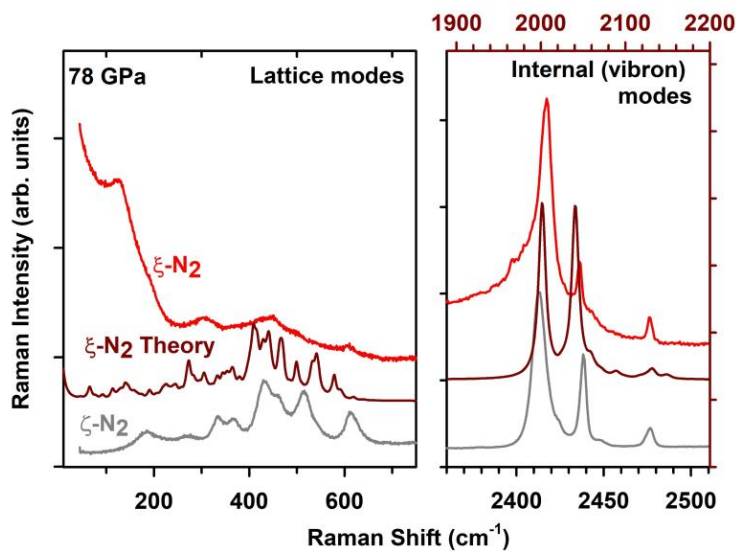

**Figure 3.** Raman spectra of ξ -$N_2$ in the heated region (Ag heat absorber) compared with ζ-$N_2$ in the unheated region of the same high-pressure chamber and the results of first-principles theoretical calculations for ξ -$N_2$ (using a 10 cm$^{-1}$ Lorentzian smearing). Left and right panels show the lattice and vibron spectral ranges, respectively. The upper frequency axis (dark red) shows the calculated vibron spectrum. Excitation wavelength: 532 nm.

Raman mapping of the Cu–N$_2$ sample after laser heating at 98 GPa, targeting tζ-N$_2$, reveals subtle differences between heated and unheated regions (Fig. S2, Supplemental Material). The unheated region shows spectra characteristic of ζ-N$_2$, whereas the heated region, where Cu nitrides formed, exhibits a superposition of nitride signals and those of tζ -N$_2$. Because of the similar molecular arrangement (Fig. 2), the Raman spectra of tζ -N$_2$ and ζ-N$_2$ are expected to be alike; however, the tripled unit cell of tζ-N$_2$ increases the number of Raman-active modes (Table S4, Supplemental Material; cf. Ref.[2]). Most additional modes are weak and fall within the same spectral range as the ζ-N$_2$ peaks (Fig. S3, Supplemental Material), so the strongest bands nearly coincide. A characteristic feature of tζ-N$_2$ is an additional low-frequency shear mode between molecular layers, although it cannot be definitively resolved here due to overlap with Cu-nitride signals. Clear differences appear at 150–350 cm$^{-1}$ and 2420–2480 cm$^{-1}$, where extra sharp peaks are observed (Fig. S2, Supplemental Material).

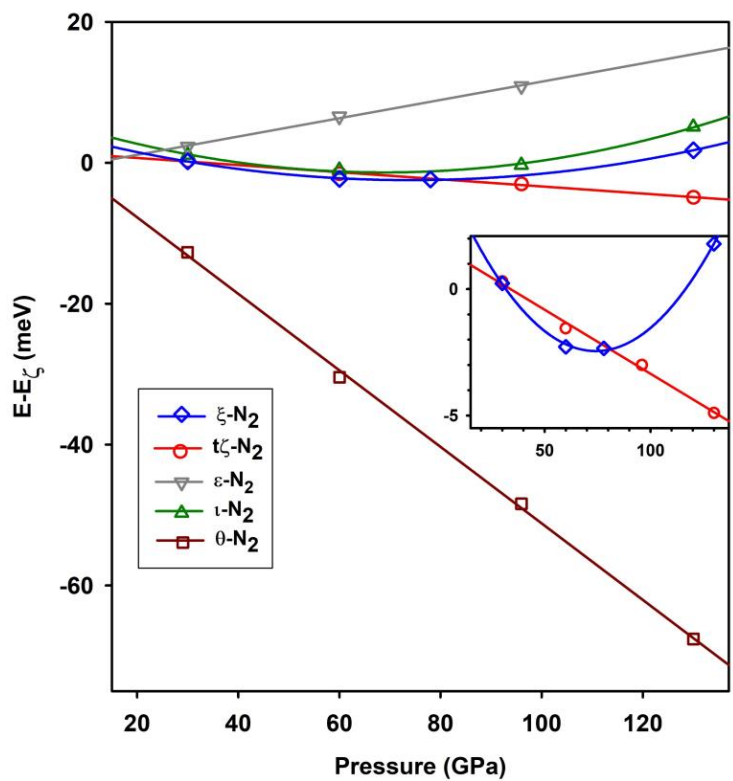

**Figure 4.** First-principles calculated enthalpies of molecular nitrogen phases as a function of pressure, referenced to ζ-N$_2$. The inset provides an expanded view for tζ-N$_2$ and ξ-N$_2$.

The discovery of two additional molecular phases of nitrogen in a system already known for its structural complexity is remarkable. First-principles calculations were performed to evaluate their stability (Fig. 4). The enthalpy-pressure curves show that the new ξ-$N_2$ and tζ-$N_2$ phases are metastable with respect to θ-$N_2$, but remain competitive to ζ-$N_2$ and ι-$N_2$ phases. In the 30-80 GPa range, ξ-$N_2$ is the second most stable molecular phase. Thus, its formation upon heating ζ-$N_2$ at 78 GPa, where ζ-$N_2$ would typically transform to θ-$N_2$ or ι-$N_2$, is plausible.

The enthalpy of tζ-$N_2$ is very close to that of ζ-$N_2$ (Fig. 4). In the 80–130 GPa range, tζ-$N_2$ is more stable than ι-$N_2$ and new ξ-$N_2$, but less stable than θ-$N_2$. Notably, above ~35 GPa tζ-$N_2$ becomes slightly more stable than ζ-$N_2$, consistent with its formation at 98 GPa after prolonged heating. Because the XRD and Raman signatures of tζ-$N_2$ differ only subtly from those of ζ-$N_2$, the transformation previously assigned to κ-$N_2$ at 110–130 GPa [5] may correspond to ζ-$N_2$ → tζ-$N_2$. The appearance of an additional low-frequency Raman mode, observed in separate experiments up to 118 GPa (Fig. S4, Supplemental Material), supports this interpretation and suggests that the transition is sluggish at room temperature but accelerated by moderate heating.

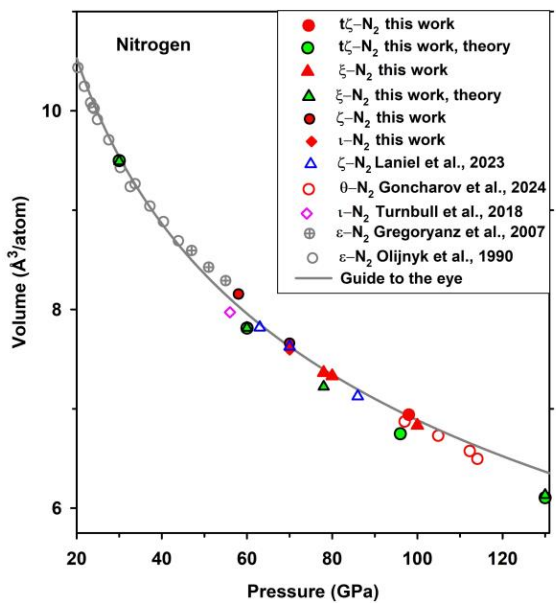

**Figure 5.** Volume–pressure data for various nitrogen phases at 300 K. Filled symbols correspond to measurements from this work; open symbols represent previous experimental results [1-3, 5, 30]. The solid line shows a provisional 300 K isotherm based on the Vinet equation of state ($P_0 = 5.0$ GPa, $V_0 = 14.6$ Å$^3$, $K_0 = 19.4$ GPa, $K_0' = 5.5$) drawn through the combined dataset.

Comparison of molecular volumes (Fig. 5) shows that $\xi$-$N_2$, $t\zeta$-$N_2$, $\iota$-$N_2$, and $\zeta$-$N_2$ follow trends reported previously, whereas $\theta$–$N_2$ are denser in the limit of high pressures, consistent with their greater stability (Fig. 4).

First-principles calculations reproduce the experimental volume trends, showing that $t\zeta$-$N_2$ denser than $\xi$-$N_2$ by ~1–2% over 30–130 GPa, consistent with its greater stability at high pressure. Furthermore, first-principles theoretical calculations of the phonon dispersion curves (Fig. S5, Supplementary Materials) demonstrate dynamic stability of the new $N_2$ phases under the pressure conditions of their synthesis.

Our experiments reveal two previously unknown molecular phases of nitrogen, $\xi$-$N_2$ and $t\zeta$-$N_2$, synthesized in laser-heated DACs at 78–98 GPa and quenched to room temperature. These structures contain a record number of molecules per unit cell—56 for $\xi$-$N_2$ and 48 for $t\zeta$-$N_2$ (equal to $\iota$-$N_2$)—demonstrating the capability of single-crystal XRD to resolve extremely complex molecular crystals at sub-100 GPa pressures. The structure of $\xi$-$N_2$ is unusual for a diatomic solid, showing elements of a host–guest architecture in which central molecules are confined within cages formed by surrounding molecules.

In contrast, $t\zeta$-$N_2$ is a polytypic modification of the common room-temperature $\zeta$-$N_2$ and is energetically competitive, exceeded in stability only by $\theta$-$N_2$ and its polytype $\lambda$-$N_2$ [7], which are difficult to access kinetically. Calculations predict $t\zeta$-$N_2$ to become slightly more stable than $\zeta$-$N_2$ above ~35 GPa, consistent with its formation at 98 GPa after prolonged heating. The subtle differences between the two phases suggest that $t\zeta$-$N_2$ may correspond to the $\kappa$-$N_2$ phase reported at higher pressures and that the $\zeta$-$N_2$ → $t\zeta$-$N_2$ transition is sluggish at room temperature but facilitated by heating.

These results demonstrate that the high-pressure phase diagram of nitrogen is exceptionally rich and likely incomplete, with multiple stable and metastable molecular structures persisting near the molecular-to-polymeric transition. Their formation is highly sensitive to kinetic conditions, enabling diverse pathways to polymeric states.

Support is acknowledged by the National Science Foundation Grant No. CHE-2302437, Grant Nos. DMR-2200670, and Carnegie Science. Portions of this work were performed at GeoSoilEnviroCARS (The University of Chicago, Sector 13), Advanced Photon Source (APS), Argonne National Laboratory. GeoSoilEnviroCARS is supported by the National Science Foundation–Earth Sciences (EAR–1634415). Portions of this work were performed at HPCAT (sector 16) of the Advanced Photon Source (APS), Argonne National Laboratory. HPCAT operations are supported by DOE-NNSA's Office of Experimental Sciences. This research used resources of the Advanced Photon Source, a U.S. Department of Energy (DOE) Office of Science User Facility operated for the DOE Office of Science by Argonne National Laboratory under Contract No. DE-AC02-06CH11357.
1. R. Turnbull, M. Hanfland, J. Binns, M. Martinez-Canales, M. Frost, M. Marqués, R. T. Howie and E. Gregoryanz, Nature Communications **9** (1), 4717 (2018).
2. D. Laniel, F. Trybel, A. Aslandukov, J. Spender, U. Ranieri, T. Fedotenko, K. Glazyrin, E. L. Bright, S. Chariton, V. B. Prakapenka, I. A. Abrikosov, L. Dubrovinsky and N. Dubrovinskaia, Nature Communications **14** (1), 6207 (2023).
3. A. F. Goncharov, I. G. Batyrev, E. Bykova, L. Brüning, H. Chen, M. F. Mahmood, A. Steele, N. Giordano, T. Fedotenko and M. Bykov, Physical Review B **109** (6), 064109 (2024).
4. E. Gregoryanz, A. F. Goncharov, R. J. Hemley, H.-k. Mao, M. Somayazulu and G. Shen, Physical Review B **66** (22), 224108 (2002).
5. E. Gregoryanz, A. F. Goncharov, C. Sanloup, M. Somayazulu, H.-k. Mao and R. J. Hemley, The Journal of Chemical Physics **126** (18), 184505 (2007).
6. D. Tomasino, Z. Jenei, W. Evans and C.-S. Yoo, The Journal of Chemical Physics **140** (24), 244510 (2014).
7. M. Frost, R. T. Howie, P. Dalladay-Simpson, A. F. Goncharov and E. Gregoryanz, Physical Review B **93** (2), 024113 (2016).
8. A. F. Goncharov, E. Gregoryanz, H.-K. Mao and R. J. Hemley, Low Temperature Physics **27** (9), 866-869 (2001).
9. R. Bini, L. Ulivi, J. Kreutz and H. J. Jodl, The Journal of Chemical Physics **112** (19), 8522-8529 (2000).
10. J. Yan, P. Dalladay-Simpson, L. J. Conway, F. Gorelli, C. Pickard, X.-D. Liu and E. Gregoryanz, Scientific Reports **14** (1), 16394 (2024).
11. M. I. Eremets, A. G. Gavriliuk, N. R. Serebryanaya, I. A. Trojan, D. A. Dzivenko, R. Boehler, H. K. Mao and R. J. Hemley, The Journal of Chemical Physics **121** (22), 11296-11300 (2004).
12. C. J. Pickard and R. J. Needs, Physical Review Letters **102** (12), 125702 (2009).
13. W. D. Mattson, D. Sanchez-Portal, S. Chiesa and R. M. Martin, Physical Review Letters **93** (12), 125501 (2004).
14. W. Sontising and G. J. O. Beran, Physical Review Materials **4** (6), 063601 (2020).
15. H. Alkhaldi and P. Kroll, The Journal of Physical Chemistry C **123** (12), 7054-7060 (2019).
16. A. Erba, L. Maschio, C. Pisani and S. Casassa, Physical Review B **84** (1), 012101 (2011).
17. D. Melicherová and R. Martoňák, The Journal of Chemical Physics **158** (24) (2023).
18. E. Gregoryanz, A. F. Goncharov, R. J. Hemley and H.-k. Mao, Physical Review B **64** (5), 052103 (2001).
19. A. F. Goncharov, E. Bykova, M. Bykov, E. Edmund, J. S. Smith, S. Chariton and V. B. Prakapenka, Physical Review Materials **7** (5), 053604 (2023).
20. M. Bykov, E. Bykova, C. J. Pickard, M. Martinez-Canales, K. Glazyrin, J. S. Smith and A. F. Goncharov, Physical Review B **104** (18), 184105 (2021).

21.	E. Bykova, I. G. Batyrev, M. Bykov, E. Edmund, S. Chariton, V. B. Prakapenka and A. F. Goncharov, Physical Review B **108** (2), 024104 (2023).
22.	H. Olijnyk and A. P. Jephcoat, Physical Review Letters **83** (2), 332-335 (1999).
23.	A. Dewaele, M. Torrent, P. Loubeyre and M. Mezouar, Physical Review B **78** (10), 104102 (2008).
24.	A. Dewaele, P. Loubeyre and M. Mezouar, Physical Review B **70** (9), 094112 (2004).
25.	V. B. Prakapenka, A. Kubo, A. Kuznetsov, A. Laskin, O. Shkurikhin, P. Dera, M. L. Rivers and S. R. Sutton, High Pressure Research **28** (3), 225-235 (2008).
26.	A. Liang, H. R. A. ten Eikelder, U. Ranieri, J. Spender, B. Massani, T. Fedotenko, K. Glazyrin, N. Giordano, E. Lawrence Bright, J. Wright, L.-T. Shi, F. Trybel and D. Laniel, JACS Au **6** (1), 193-199 (2026).
27.	J. Binns, M.-E. Donnelly, M. Peña-Alvarez, M. Wang, E. Gregoryanz, A. Hermann, P. Dalladay-Simpson and R. T. Howie, The Journal of Physical Chemistry Letters **10** (5), 1109-1114 (2019).
28.	W. Yi, K. Zhao, Z. Wang, B. Yang, Z. Liu and X. Liu, ACS Omega **5** (11), 6221-6227 (2020).
29.	W. Yi, Y. Zhang, G. Zhang and X. Liu, Physical Chemistry Chemical Physics **27** (11), 5902-5908 (2025).
30.	H. Olijnyk, The Journal of Chemical Physics **93** (12), 8968-8972 (1990).

# Single-crystal XRD measurements

Single-crystal X-ray diffraction (SCXRD) was performed at beamlines 16-ID-B (HPCAT) and 13-ID-CD (GSECARS) of the Advanced Photon Source (APS), Argonne National Laboratory. Monochromatic X-ray beams were focused to 1–3 µm spots, with wavelengths of 0.34453 Å (HPCAT) and 0.3344 and 0.4133 Å (GSECARS). Diffraction was recorded using area detectors: PILATUS3 X 2M CdTe (HPCAT) and EIGER2 S CdTe 9M (GSECARS). For the single-crystal XRD measurements, samples were rotated around a vertical ω-axis in a range of ±30° with an angular step $\Delta\omega$ = 0.5° and an exposure time of 1-2 s/frame. For analysis of the single-crystal diffraction data we used the CrysAlisPro software package [1], which facilitates the SCXRD analysis of multigrain samples. To calibrate an instrumental model in the CrysAlisPro software, i.e., the sample-to-detector distance, detector's origin, offsets of goniometer angles, and rotation of both X-ray beam and the detector around the instrument axis, we used a single crystal of orthoenstatite ($(Mg_{1.93}Fe_{0.06})(Si_{1.93}, Al_{0.06})O_6$, *Pbca* space group, *a* = 8.8117(2), *b* = 5.1832(1), and *c* = 18.2391(3) Å). The structure was solved with the ShelXT structure solution program and refined with the Olex2 program[2, 3].

# First-principles theoretical calculations

First-principles calculations were performed within the framework of density functional theory (DFT), implemented in CASTEP code [4]. For the exchange-correlation functional, we employed the generalized gradient approximation (GGA) with the Perdew-Burke-Ernzerhof (PBE)[5] parameterization. Plane-wave basis with a cut-off energy of 700 eV and dense k-point grid were used to calculate total energy. We relaxed the crystal structure until all the stress components and forces of atoms were smaller than 0.01 eV/ Å. First-principles theoretical calculations have been performed in *P*4$_1$2$_1$2 (θ), *R$\bar{3}$c* (ε), *C*2/c (ζ), *P*2$_1$/*c* (ι), *P*6*cc* (ξ), and *C*2/*c* (tζ) phases at selected pressures (30, 60, 96, and 130 GPa), where these structures were optimized using norm-conserving pseudopotentials. To approximate the disordered ξ-N$_2$, we model it using the ordered *P2* structure. Monkhorst-Pack grid providing 0.025 Å$^{-1}$ separation for accurate k-points sampling of the Brillouin Zone (BZ) is used for all the structures [5]. The phonon dispersion and phonon frequency calculations were performed using linear response and finite displacement methods implemented in the CASTEP code[6]. The Raman spectra were calculated using the formalism presented in Ref. [7]. The electronic bandgap calculations have been performed within GGA/PBE approximations.

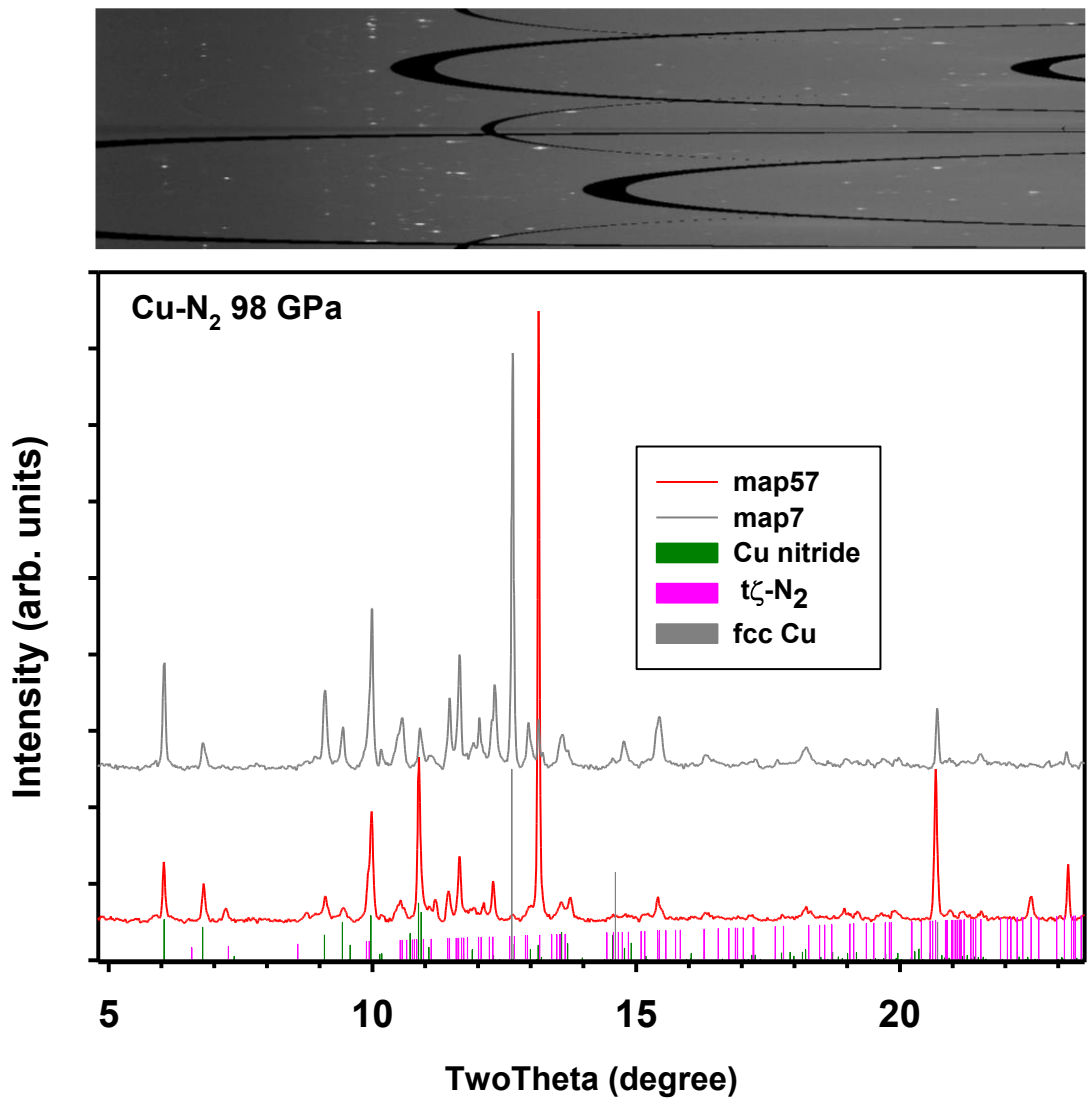

**Figure S1.** X-ray-diffraction (XRD) patterns of Cu-$N_2$ samples after laser heating at 98 GPa. Main panel (bottom) shows integrated 1D XRD curves measured at different sample positions (maps 7 and map 57). Vertical bars indicate the Bragg peaks of the fitted phases. The top curve (shifted for clarity) is well represented by a mixture of diffraction patterns of a new Cu nitride (the structure and composition will be presented elsewhere) and fcc Cu. The bottom curve can be understood as a superposition of the XRD signal of the same nitride and a new ζ-$N_2$ with a triple unit cell— tζ-$N_2$. Top panel shows a 2D image of the diffractogram in the position 57 in rectangular coordinates, corresponding to the bottom curve in the main panel, where tζ-$N_2$ is present. The x-ray wavelength is 0.4133 Å.

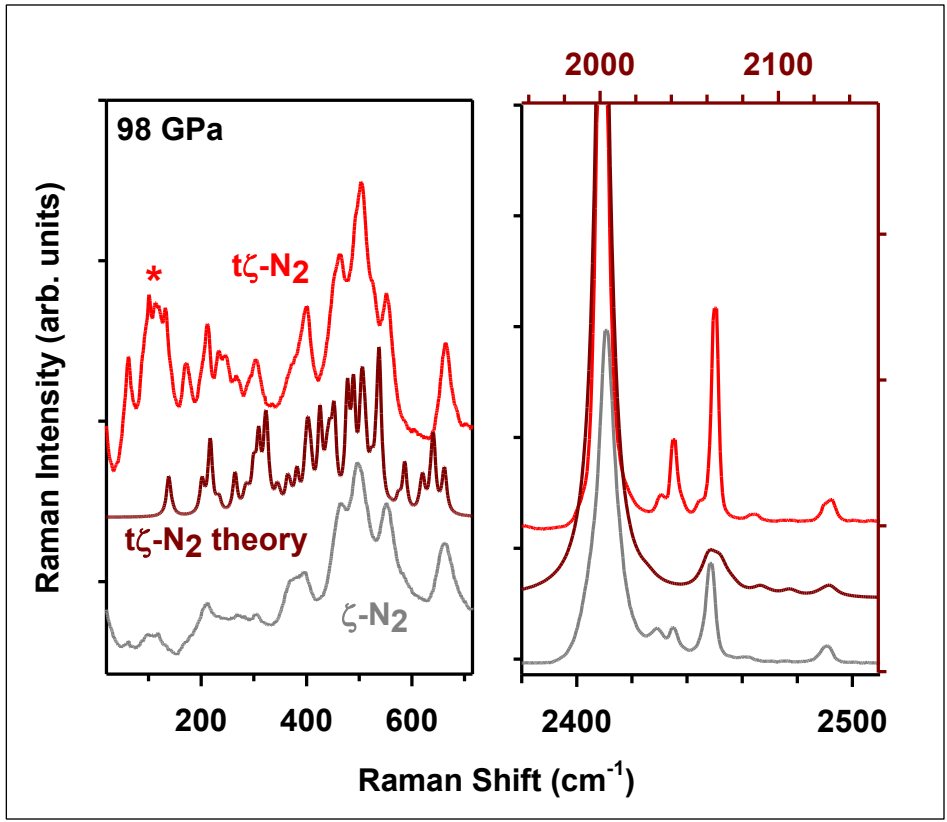

**Figure S2.** Raman spectra of tζ-N$_2$ in the heated area (Cu heat absorber) compared to spectrum of ζ-N$_2$ in the unheated area of the same high-pressure cavity and the results of first-principles theoretical calculations for tζ-N$_2$ (using a 10 cm$^{-1}$ Lorentzian smearing). The panels correspond to the spectral ranges of the lattice and vibron modes. The asterisk in the left panel marks the bands assigned to Cu nitride. The top frequency axis (dark red) corresponds to the theoretically calculated spectrum of vibron modes. The excitation wavelength is 660 nm.

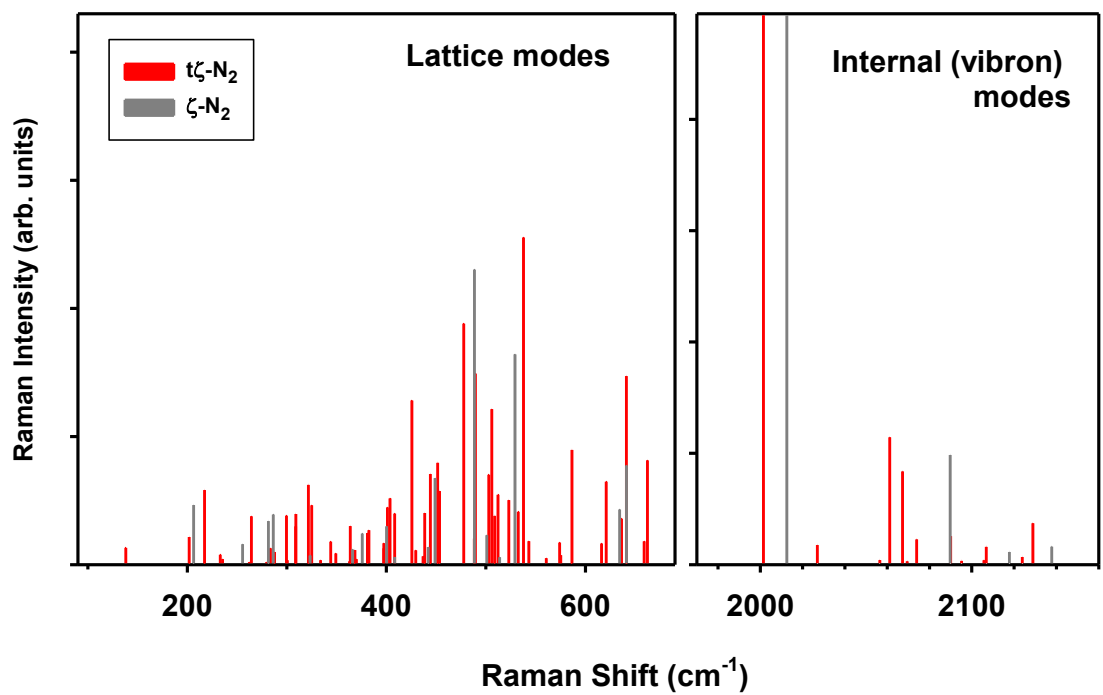

**Figure S3.** Comparison of the computed Raman spectra of tζ-N$_2$ and ζ-N$_2$ phases at 98 GPa.

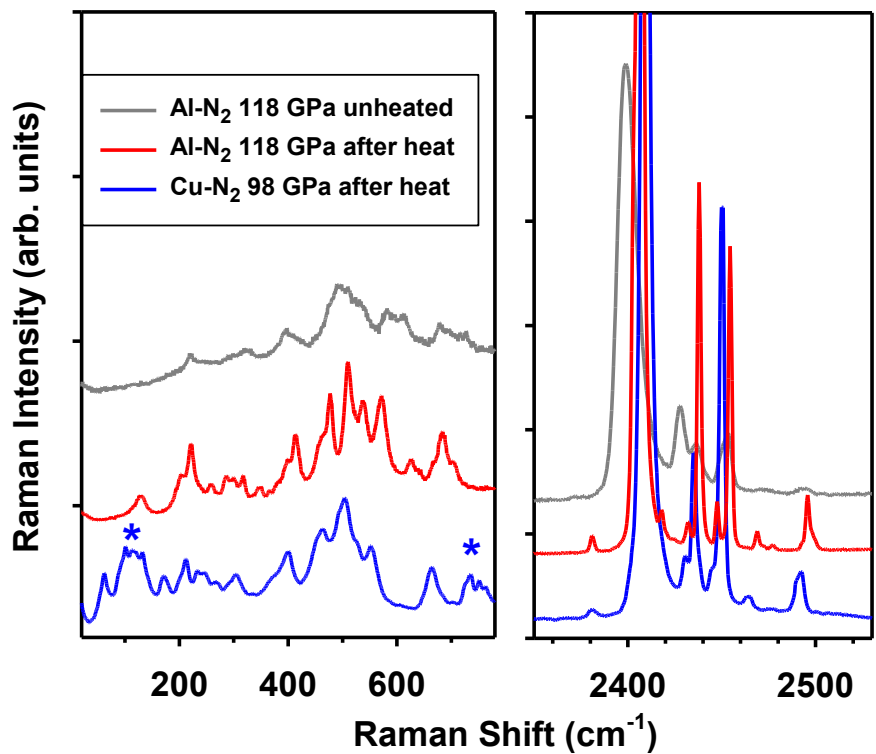

**Figure S4.** Raman spectra of tζ-$N_2$ in the heated area (Cu and Al heat absorbers) compared to spectrum of ζ-$N_2$ in the unheated area of the same high-pressure cavity. The panels correspond to the spectral ranges of the lattice and vibron modes. The asterisks in the left panel mark bands assigned to Cu nitride. The excitation wavelength is 660 nm (Cu-$N_2$ experiment) and 532 nm (Al-$N_2$ experiment).

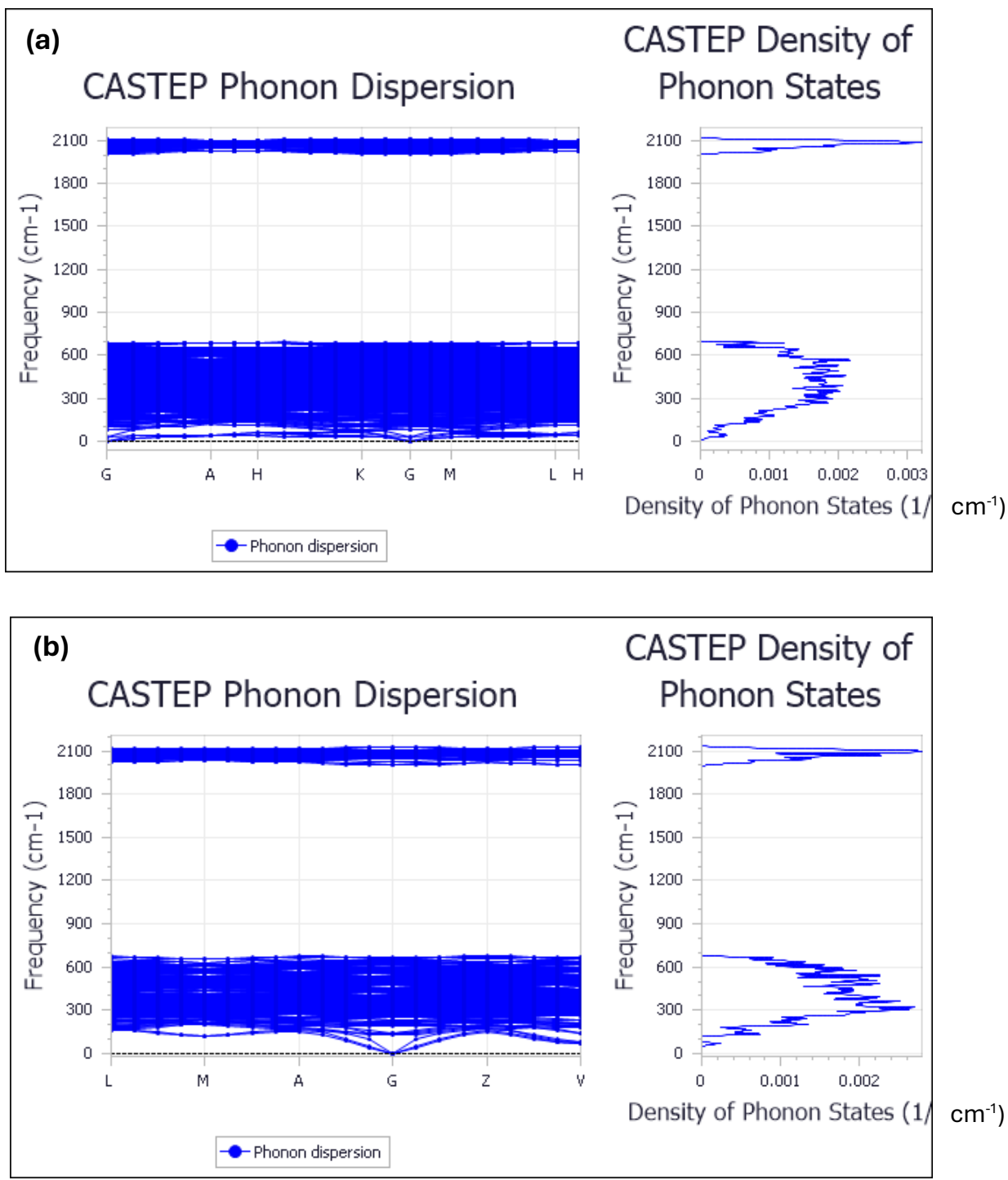

**Figure S5.** Theoretically calculated phonon dispersion curves of ξ-N$_2$ (a) and tζ-N$_2$ (b) at 78 and 98 GPa, respectively.

**Table S1.** Experimental observations in the Ag-$N_2$ system.

| Experiment ID | XRD experiment location | Pressure (GPa) | Laser heating duration (s) | Estimated temperature (K) | $N_2$ transition /structure | Reaction product |
|---|---|---|---|---|---|---|
| CC278 | HPCAT | 58 | 10 | 1600 | ε-$N_2$ →ζ-$N_2$ | None |
| CC278 | HPCAT | 78 | 10 | 1800 | ζ-$N_2$→ξ-$N_2$ | None |
| CC278 | GSECARS | 80 | 600 | 1400-2000 | ζ-$N_2$ | $AgN_5$ |
| BX90-1 | GSECARS | 70 | 900 | 1300-1700 | ζ-$N_2$→ι-$N_2$ | None |
| BX90-1 | GSECARS | 106 | 240 | 1400-3000 | ζ-$N_2$ | $AgN_5$ |

**Single-crystal XRD data for ξ-$N_2$**

Single-crystal X-ray diffraction data were collected at 293 K using synchrotron radiation with a wavelength of 0.34453 Å. The structure was solved by direct methods and refined by full-matrix least-squares minimization on $F^2$. The compound crystallizes in the hexagonal space group *P*6*cc* with lattice parameters a = 14.105(3) Å and c = 4.786(1) Å. The final agreement factors were $R_1$ = 0. 0872 for reflections with I > 2σ(I) and $wR_2$ = 0.2325 for all data, with a goodness-of-fit of 1.051. The relatively high residuals are attributed to the light-element composition, possible partial occupancies, and the use of short-wavelength synchrotron radiation. The Cambridge Structural Database[8] contains the supplementary crystallographic data for this work. These data can be obtained free of charge from FIZ Karlsruhe [9].

**Table S2. Crystallographic data and structure refinement parameters for ξ-$N_2$ at 78 GPa**

| Parameter | Value |
| --- | --- |
| Empirical formula | $N_2$ |
| Formula weight (g·mol$^{-1}$) | 21.02 |
| Temperature (K) | 293(2) |
| Radiation type | Synchrotron |
| Wavelength (Å) | 0.34453 |
| Crystal system | Hexagonal |
| Space group | *P*6*cc* |
| *a* (Å) | 14.105(3) |
| *b* (Å) | 14.105(3) |
| *c* (Å) | 4.786(1) |
| α (°) | 90 |
| β (°) | 90 |
| γ (°) | 120 |
| Unit cell volume (Å$^3$) | 824.7(4) |
| *Z* | 112 |
| Calculated density (g·cm$^{-3}$) | 3.16 |
| Absorption coefficient (mm$^{-1}$) | 0.081 |
| θ range for data collection (°) | 2.9640 – 15.7020 |
| Reflections collected | 950 |
| Independent reflections | 570 |
| Absorption correction | None |

| Parameter | Value |
| --- | --- |
| Refinement method | Full-matrix least squares on $F^2$ |
| $R_1$ [$I > 2\sigma(I)$] | 0. 0872 |
| $wR_2$ (all data) | 0.2325 |
| Goodness-of-fit on $F^2$ | 1.051 |

**Fractional atomic coordinates (*x, y, z*) and equivalent isotropic displacement parameters $U_{eq}$ (Å$^2$).**

| Atom | x | y | z | $U_{eq}$ (Å$^2$) | Occ. |
| --- | --- | --- | --- | --- | --- |
| N001 | 0.5409(6) | 0.5077(6) | 0.7869(15) | 0.0212(13) | 1.0 |
| N002 | 0.7529(11) | 0.9198(5) | 0.4042(14) | 0.0207(14) | 1.0 |
| N003 | 0.8160(6) | 0.9507(5) | 0.5598(15) | 0.0237(15) | 1.0 |
| N004 | 0.6802(6) | 0.5475(5) | 0.4769(16) | 0.0187(14) | 1.0 |
| N005 | 0.7551(5) | 0.5823(6) | 0.5987(18) | 0.0233(15) | 1.0 |
| N006 | 0.7553(4) | 0.4291(5) | 0.3950(15) | 0.0185(13) | 1.0 |
| N007 | 0.8151(5) | 0.4479(5) | 0.5567(13) | 0.0201(13) | 1.0 |
| N008 | 0.9511(6) | 0.6285(6) | 0.5760(15) | 0.0227(13) | 1.0 |
| N009 | 0.9349(5) | 0.6725(5) | 0.4254(11) | 0.0214(15) | 1.0 |
| N010 | 0.979(3) | 0.9536(8) | 0.328(3) | 0.004(3) | 0.333 |

**Single-crystal XRD data for tζ-N$_2$ at 98 GPa**

Single-crystal X-ray diffraction data were collected at 293 K using synchrotron radiation with a wavelength of 0.4133 Å. The structure was solved by direct methods and refined by full-matrix least-squares refinement on F$^2$. The compound crystallizes in the monoclinic space group *C2/c* with lattice parameters a = 7.1009(17) Å, b = 6.5202(12) Å, c = 14.614(7) Å, and β = 100.05(4)°. The final refinement converged with R$_1$ = 0.0785 for reflections with I > 2σ(I) and wR$_2$ = 0.2288 for all data. The Cambridge Structural Database [8] contains the supplementary crystallographic data for this work. These data can be obtained free of charge from FIZ Karlsruhe [9].

**Table S3. Crystallographic data and structure refinement parameters for tζ-N$_2$ at 98 GPa**

| Parameter | Value |
| --- | --- |
| Empirical formula | N$_2$ |
| Formula weight (g·mol$^{-1}$) | 28.01 |
| Temperature (K) | 293(2) |
| Radiation type | Synchrotron |
| Wavelength (Å) | 0.4133 |
| Crystal system | Monoclinic |
| Space group | *C2/c* |
| a (Å) | 7.1009(17) |
| b (Å) | 6.5202(12) |
| c (Å) | 14.614(7) |
| α (°) | 90 |
| β (°) | 100.05(4) |
| γ (°) | 90 |

| Parameter | Value |
|---|---|
| Unit cell volume (Å$^3$) | 666.2(4) |
| Z | 96 |
| Calculated density (g·cm$^{-3}$) | 3.352 |
| Absorption coefficient (mm$^{-1}$) | 0.103 |
| Reflections collected | 442 |
| Independent reflections | 373 |
| $R_{int}$ | 0.0162 |
| Absorption correction | Not applied |
| Refinement method | Full-matrix least squares on $F^2$ |
| $R_1$ [$I > 2\sigma(I)$] | 0.0785 |
| $wR_2$ (all data) | 0.2288 |
| Goodness-of-fit on $F^2$ | 1.034 |

**Fractional atomic coordinates (*x, y, z*) and equivalent isotropic displacement parameters $U_{eq}$ (Å$^2$) for t$\zeta$-N$_2$ at 98 GPa.**

| Atom | x | y | z | $U_{eq}$ (Å$^2$) | Occ. |
|---|---|---|---|---|---|
| N1 | 0.2046(6) | 0.2781(5) | 0.0218(3) | 0.0198(10) | 1.0 |
| N2 | 0.0750(6) | 0.4656(5) | 0.7501(4) | 0.0190(10) | 1.0 |
| N3 | 0.1520(6) | 0.1017(5) | 0.1369(3) | 0.0195(9) | 1.0 |
| N4 | 0.0905(5) | 0.1954(5) | 0.1847(3) | 0.0182(9) | 1.0 |
| N5 | 0.1361(5) | 0.0812(5) | 0.4734(3) | 0.0193(10) | 1.0 |
| N6 | 0.0844(6) | 0.1902(5) | 0.5194(3) | 0.0190(10) | 1.0 |

| Atom | x | y | z | $U_{eq}$ (Å$^2$) | Occ. |
|---|---|---|---|---|---|
| N7 | 0.2852(6) | 0.2900(5) | 0.3068(3) | 0.0221(10) | 1.0 |
| N8 | 0.2047(6) | 0.2490(5) | 0.3581(3) | 0.0206(10) | 1.0 |
| N9 | 0.0795(5) | 0.4792(4) | 0.4266(3) | 0.0182(10) | 1.0 |
| N10 | −0.0711(6) | 0.5000(4) | 0.4037(3) | 0.0197(9) | 1.0 |
| N11 | 0.1297(5) | 0.0618(5) | 0.8086(3) | 0.0183(10) | 1.0 |
| N12 | 0.0837(5) | 0.1779(5) | 0.8526(3) | 0.0176(9) | 1.0 |

**Table S4. Vibrational modes and their Raman and IR activity in *P3c1* ($C_{3v}^3$) and *C2/c* ($C_{2h}^6$) crystal structures**

| Space group | ξ-$N_2$ *P6cc* #184 ($C_{6v}^2$) | | tξ-$N_2$ *C2/c* #15 ($C_{2h}^6$) | |
|---|---|---|---|---|
| Number of molecules in the unit cell | 56 | | 48 | |
| Site symmetry | 12d ($C_1$) | | 8f ($C_1$) | |
| Acoustic modes | $A_1+E_1$ | | $A_u+2B_u$ | |
| Optical modes | Modes | Activity | Modes | Activity |
| Total Optical Modes | $27A_1+55E_1$ +$56E_2$ +$28A_2+28B_1+28B_2$ | Raman+IR Raman inactive | $36A_g+36B_g+$ $35A_u+34B_u$ | Raman IR |
| Intramolecular | $6A_1+9E_1$ +$9E_2$ +$6A_2+4B_1+4B_2$ | Raman+IR Raman inactive | $6A_g+6B_g+$ $6A_u+6B_u$ | Raman IR |
| Translational | $13A_1+27E_1$ +$28E_2$ +$14A_2+14B_1+14B_2$ | Raman+IR Raman inactive | $18A_g+18B_g+$ $17A_u+16B_u$ | Raman IR |
| Librational | $8A_1+19E_1$ +$19E_2$ +$8A_2+10B_1+10B_2$ | Raman+IR Raman inactive | $12A_g+12B_g+$ $12A_u+12B_u$ | Raman IR |

1. O. CrysAlisPro Software System (Rigaku Oxford Diffraction) (Oxford, UK, 2014).
2. G. Sheldrick, Acta Crystallographica Section A **71** (1), 3-8 (2015).
3. O. V. Dolomanov, L. J. Bourhis, R. J. Gildea, J. A. K. Howard and H. Puschmann, Journal of Applied Crystallography **42** (2), 339-341 (2009).
4. G. Kresse and J. Furthmüller, Physical Review B **54** (16), 11169-11186 (1996).
5. H. J. Monkhorst and J. D. Pack, Physical Review B **13** (12), 5188-5192 (1976).
6. K. Refson, P. R. Tulip and S. J. Clark, Physical Review B **73** (15), 155114 (2006).
7. D. Porezag and M. R. Pederson, Physical Review B **54** (11), 7830-7836 (1996).
8. C. R. Groom, I. J. Bruno, M. P. Lightfoot and S. C. Ward, Acta Crystallographica Section B **72** (2), 171-179 (2016).
9. www.ccdc.cam.ac.uk/structures.